\documentclass[12pt]{article}
\usepackage{amsmath,amsfonts}
\advance \textheight by \topskip
\usepackage{cite}
\usepackage{filecontents}

\makeatletter
\@addtoreset{equation}{section}
 
\makeatother

\newcommand{\be}{\begin{equation}}
\newcommand{\ee}{\end{equation}}
\newcommand{\bea}{\begin{eqnarray}}
\newcommand{\eea}{\end{eqnarray}}
\newcommand{\dd}{\partial}

\def\>{\rangle}
\def\<{\langle}

\begin{document}

\title{
{\bf On some integrable deformations of the  Wess-Zumino-Witten model}}

\author{
{\sf   N. Mohammedi} \thanks{e-mail:
noureddine.mohammedi@univ-tours.fr}$\,\,$${}$
\\
{\small ${}${\it Institut  Denis Poisson (CNRS - UMR 7013), }} \\
{\small {\it Universit\'e  de Tours,}}\\
{\small {\it Facult\'e des Sciences et Techniques,}}\\
{\small {\it Parc de Grandmont, F-37200 Tours, France.}}}

\date{}
\maketitle
\vskip-1.5cm

\vspace{2truecm}

\begin{abstract}

\noindent
Lie algebra valued equations translating the integrability of a general 
two-dimensional Wess-Zumino-Witten model are  given.
We found a simple solution to these equations and identified a new
integrable  non-linear sigma model. This is a two-parameter deformation
of the  Wess-Zumino-Witten model.

\end{abstract}

\newpage

\setcounter{equation}{0}

\section{Introduction}

\par
The search for integrable two dimensional non-linear sigma model has known various 
developments. The early attempts dealt mostly with deformations of the principal chiral sigma model 
and examples based on the Lie algebra $SU(2)$ were found \cite{cherednik,hlavaty}. Later other integrable
Wess-Zumino-Witten models, involving the Lie algebra $SU(2)$, were constructed \cite{forgacs,yoshida1,yoshida2}.
The revival of the subject came after the work of Klim\v{c}\'{\i}k  on the so-called Yang-Baxter deformation of the
principal chiral model \cite{klimcik}.  More recently Sfetsos presented a method for constructing integrable deformation
of the Wess-Zumino-Witten model \cite{sfetsos}. Various issues were treated later in the literature 
\cite{a,b,bb,bbb,c,cc,d,e,ee,f,g,gg,ggg,h,i,j,k,l,m,n,o,p,q,u}
and a nice account of these can be found in \cite{seibold} and references within.
Our interest in integrable non-linear sigma models is motivated by their relation to string theories \cite{gsw}.
The hope is to find more solvable string theories and their spectrum in non-trivial backgrounds along
the lines in \cite{tseytlin1,tseytlin2, tseytlin3}.   
\par
In \cite{I1,I2} we have given the conditions for the most general non-linear sigma model to be integrable.
These were specified in terms of the geometry and the structure of the target space manifold.
A general two-dimensional non-linear sigma model is given  by the action{\footnote{The two-dimensional coordinates 
are
$\left(\tau,\sigma\right)$ with $\dd_0=\frac{\dd}{\dd\tau}$ and $\dd_1=\frac{\dd}{\dd\sigma}$.
In the rest of the paper, however, we will use the complex coordinates
$\left(z=\tau +i\sigma\,,\,\bar z=\tau -i\sigma\right)$ together with
$\dd= \frac{\dd}{\dd z}$ and $\bar\dd=\frac{\dd}{\dd\bar z}$. Our conventions are such that the alternating tensor is 
$\epsilon^{z\bar{z}}=+1$.}}
\be
S=\int{\rm d}z{\rm d}\bar z \left[G_{ij}\left(\varphi\right)+
B_{ij}\left(\varphi\right)\right]
\dd\varphi^i\bar\dd\varphi^j\,\,\,.
\label{sigma-1}
\ee
The invertible metric $G_{ij}$ and the anti-symmetric tensor $B_{ij}$ are the backgrounds of the bosonic string theory.
The equations of motion of this theory are
\be
\bar\dd\dd\varphi^l +\Omega^l_{ij}
\dd\varphi^i
\bar\dd\varphi^j=0\,\,\,\,,\,\,\,\, \Omega^k_{ij} =  \Gamma^k_{ij} - H^k_{ij}\,\,\,\,\,,
\label{eqofmot}
\ee
where $\Gamma^k_{ij} $ and  $H^k_{ij} = \frac{1}{2}G^{kl}
\left(\dd_lB_{ij}+\dd_jB_{li}+\dd_iB_{jl}\right)$ are, respectively, the Christoffel
symbols and the torsion. 

\par
The equations of motion can be cast, for all values of the parameter $\mu$,  in the form of a zero curvature relation
\be
\left[\dd +\frac{1}{1+\mu}\left(K_i-L_i\right)\dd\varphi^i\,\,,\,\,
\bar\dd +\frac{1}{1-\mu}\left(K_j+L_j\right)\bar\dd\varphi^i
 \right]\,=\,0
\label{zero-curv}
\ee
if the space manifold is equipped with two sets of matrices $K_i(\varphi)$ and  $L_i(\varphi)$
satisfying 
\bea
&&\dd_i K_j +\dd_j K_i -2\Gamma^l_{ij}\,K_l=0\,\,\,\,\,,
\nonumber
\\
&&\dd_i L_j -\dd_j L_i + 2H^l_{ij}\,K_l = 0 \,\,\,\,\,,
\nonumber 
\\
&&\dd_i L_j +\dd_j L_i -2\Gamma^l_{ij}\,L_l=
 \left[L_i\,,\,K_j\right]+\left[L_j\,,\,K_i\right]\,\,\,\,\,,
\nonumber
\\
&& \dd_i K_j -\dd_j K_i + 2H^l_{ij}\,L_l
= 
\left[L_i\,,\,L_j\right]-\left[K_i\,,\,K_j\right]\,\,\,\,\,.
\label{combination-1} 
\eea
The last two equations determine the structure of the space manifold of the non-linear sigma model.
On the other hand, the first two relations indicate that the non-linear sigma model is symmetric 
under a global isometry transformation \cite{jack,hull} with  
$J=\left(K_i-L_i\right)\dd\varphi^i$ and  $\bar J=\left(K_i+L_i\right)\bar\dd\varphi^i$
being the conserved currents. The zero curvature relation is then the same as the
two equations $\dd \bar J + \bar \dd  J=0$ and  $\dd \bar J - \bar \dd  J +\left[J\,,\,\bar J\right]=0$.

\par 
Athough the conditions (\ref{combination-1}) specify the geometry of the manifold \cite{I2}, their general solutions are not 
yet known. 
In this note, we 
continue this program and consider 
simpler non-linear sigma models. Namely, 
the most general integrable deformation of the Wess-Zumino-Witten (WZW) model. 
The conditions (\ref{combination-1}) are now more tractable.
They are in the form of  a Lie algebra valued relation which generalises the Yang-Baxter equation used in   \cite{klimcik}
and the integrable deformations of the principal chiral model \cite{sochen}.
We are able to find a  solutions to this integrability condition. This leads to an integrable two-dimensional non-linear sigma
model in the form of a two-parameter family of  integrable
deformations
of the Wess-Zumino-Witten model. Our result
might be a generalisation of the two-parameter integrable
deformations of the WZW model found in \cite{delduc}. Indeed, the two constructions coincide
for a special case and we conjecture that our work contains more integrable models.

\par
The paper is organised as follow: In the next section  we give in details the steps
leading to the equivalent relation to  (\ref{combination-1})  for the case 
of the general Wess-Zumino-Witten model with a  summary of  the results at the end.
For completeness, we show in section 3 how the Yang-Baxter integrable sigma model is
obtained as a particular case of our construction. In section 4, we construct the  solution
to the integrability conditions and give, in section 5,  the corresponding integrable non-linear sigma models.

\section{The general construction}

We consider the two-dimensional non-linear sigma model as defined 
by the action
\bea
S\left(g\right) &=& \int_{\dd{\cal{M}}}{\rm d}z{\rm d}\bar z \,
<g^{-1}\dd g\,\,,\,\,
\left(M+N\right)\,g^{-1}\bar\dd g>_{{\cal G}}
\nonumber \\
&+& \frac{\lambda}{6}\int_{{\cal{M}}}{\rm d}^3x\,\epsilon^{\mu\nu\rho}\,<g^{-1}\dd_\mu g\,\,,\,\,
\left[\,g^{-1}\dd_\nu g\,\,,\,\, g^{-1}\dd_\rho g\right]\,>_{{\cal G}}
\,\,\,\,,
\label{action}
\eea
where ${\cal M}$ is a  three-dimensional ball having $x^\mu$, with $\mu=1,2,3$, as coordinates
and $\dd {\cal M}$ is the boundary of this ball with coordinates $z$ and $\bar z$.
The bi-linear form $<\,,\,>_{{\cal G}}$ is the Killing-Cartan form on the Lie algebra ${\cal G}$
and the field $g(z\,,\,\bar z)$ is an element of the Lie group corresponding to ${\cal G}$. The Lie algebra
is of dimension $n$.
The Wess-Zumino-Witten term comes with a parameter $\lambda$.

\par
The  Lie algebra ${\cal{G}}$ is defined by the commutation
relations $\left[T_a\,,\,T_b\right]=f^c_{ab}T_c$.
{}For a semi-simple Lie algebra the Killing-Cartan form
is $\eta_{ab}=f_{ac}^d f_{bd}^c$ and we have
$<T_a\,,\,T_b>_{{\cal G}}=\eta_{ab}={\rm Tr}\left(T_aT_b\right)$.
However, for a non semi-simple Lie algebra the bi-linear form is such that
$<T_a\,,\,T_b>_{{\cal G}}=\eta_{ab}$ with $\eta_{ab}$ an invertible matrix 
satisfying $\eta_{ab}f^b_{cd}+\eta_{cb}f^b_{ad}=0$.

\par

The two quantities $M$ and $N$ are linear
operator acting on the generators of the  Lie algebra ${\cal G}$.
They are required to satisfy the relation
\be
<X\,,\,\left(M+N\right)Y>_{{\cal G}} \,\,=\,\, <\left(M-N\right)X\,,\,Y>_{{\cal G}}\,\,\,\,
\label{bi-linear}
\ee
for any two elements $X$ and $Y$ in the Lie algebra ${\cal G}$. In other words, $M$ is 
symmetric while $N$ is anti-symmetric with respect to $<\,,\,>_{{\cal G}}$ .

\par
Putting indices, the action of $M$ and $N$ on the generators $\left\{T_a\right\}$ of the Lie algebra 
${\cal G}$ is $M\,T_a= M_a^b\, T_b$ and  $N\,T_a= N_a^b\, T_b$ and (\ref{bi-linear}) is equivalent to
\bea
\eta_{ac}\,M^c_b &=& \eta_{bc}\,M^c_a\,\,\,\,\,,
\nonumber \\
\eta_{ac}\,N^c_b &=& - \eta_{bc}\,N^c_a\,\,\,\,\,,
\eea
where $\eta_{ab}$ is the bi-linear form corresponding to the Lie algebra ${\cal G}$
as stated above.

\par
It is useful to introduce the two quantities
\bea
A &=& g^{-1}\dd g \,\,\,\,\,\,\,,\,\,\,\,\,\,\,
\nonumber \\
\bar A &=&  g^{-1}\bar\dd g \,\,\,\,.
\label{gauge-conn}
\eea
In terms of $A$ and $\bar A$, the equations of motion of the model take the form
\bea
&& \dd\left[\left(M+N+\lambda I\right)\bar A\right]+
\bar\dd\left[\left(M-N-\lambda I\right) A\right]
\nonumber \\
&&+\left[A\,,\,\left(M+N\right)\bar A\right] 
+\left[\bar A\,,\,\left(M-N\right)A\right]=0\,\,\,\,.
\eea
Multiplying this equation by $g$ on the left and $g^{-1}$ on the right, we get
the conservation equation
\bea
\dd \bar J + \bar\dd J=0\,\,\,\,\,,
\label{cons}
\eea
where we have defined the two currents $J$ and $\bar J$ as
\bea
J &=& g\left(P^{-1}A\right)g^{-1}\,\,\,\,,
\nonumber \\
\bar J &=& g\left(Q^{-1} \bar A\right)g^{-1}\,\,\,\,.
\label{curr}
\eea
Here the two linear operators $P^{-1}$ and $Q^{-1}$, acting on $A$ and $\bar A$ only, are defined as 
\bea
P^{-1} &=& M-\left(N-\lambda I\right) \,\,\,\,,
\nonumber \\
Q^{-1} &=& M+\left(N-\lambda I\right) \,\,\,\,,
\label{PQ--1}
\eea
where $I$ is the identity operator on the elements of the  Lie algebra ${\cal G}$.

\par

The conservation equation (\ref{cons}) is a result of  to the global symmetry of the action (\ref{action}) 
under the left multiplication 
\be
g\longrightarrow hg \,\,\,\,\,\,,
\ee
where $h$ is a constant group element.

\par
It is, of course,  assumed that the two linear operators $P$ and $Q$ are invertible. Hence, the inversion of (\ref{curr})
gives
\bea
A &=& P\left(g^{-1}Jg\right)\,\,\,\,,
\nonumber \\
\bar A &=& Q\left(g^{-1}\bar J g\right)\,\,\,\,.
\label{inv-curr}
\eea
However, the two currents $A$ and $\bar A$ satisfy the Cartan-Maurer identity
\be
\dd \bar A - \bar \dd A +\left[A\,,\,\bar A\right]=0\,\,\,\,.
\label{cartan}
\ee
In terms of the currents $J$ and $\bar J$, after a use of (\ref{inv-curr}) and (\ref{cartan}), one finds
the identity
\bea
&\frac{1}{2}\left(Q-P\right)\left[g^{-1}\left(\dd \bar J +\bar\dd J\right)g\right]   &
\nonumber \\
&+\frac{1}{2}\left(Q+P\right)\left[g^{-1}\left(\dd \bar J -\bar\dd J 
+ \varepsilon\left[J\,,\,\bar J\right]\right)g\right] &
\nonumber \\
&-\frac{\varepsilon}{2}\left(Q+P\right)\left[g^{-1}Jg \,,\,g^{-1}\bar J g\right]
- Q\left[P\left(g^{-1}Jg\right) \,,\,g^{-1}\bar J g\right] &
\nonumber \\
&+ P\left[Q\left(g^{-1}\bar J g\right) \,,\,g^{-1} J g\right]
+\left[P\left(g^{-1} J g\right) \,,\, Q\left(g^{-1} \bar J g\right)\right]=0
\,\,\,\,.&
\label{bianchi}
\eea
We have added and substracted the term proportional to the constant $\varepsilon$.
At this stage  $\varepsilon$ is just a bookkeeping device but will 
later join the constant $\lambda$ to form one of the deformation parameters   $\lambda\varepsilon$. 

\par
In order to have an identity that is suitable for the concept of integrability, we demand that
the linear operators $P$ and $Q$ are such that the last four terms in (\ref{bianchi})
vanish. That is, 
\bea
&-\frac{\varepsilon}{2}\left(Q+P\right)\left[g^{-1}Jg \,,\,g^{-1}\bar J g\right]
- Q\left[P\left(g^{-1}Jg\right) \,,\,g^{-1}\bar J g\right] &
\nonumber \\
&+ P\left[Q\left(g^{-1}\bar J g\right) \,,\,g^{-1} J g\right]
+\left[P\left(g^{-1} J g\right) \,,\, Q\left(g^{-1} \bar J g\right)\right]=0
\,\,\,\,.&
\eea
Since the quantities $g^{-1}Jg$ and $g^{-1}\bar Jg$ take values in the Lie algebra ${\cal G}$,
this last equation is equivalent to requiring that
\bea
\left[PX \,,\, Q Y\right] - P\left[X \,,\, Q Y\right]- Q\left[PX \,,\,  Y\right]
=\frac{\varepsilon}{2}\left(P+Q\right)\left[X \,,\, Y\right]
\label{gen-R}
\eea
for any two Lie algebra elements $X$ and $Y$. Notice that the constant $\varepsilon$
can be absorbed by a rescaling of the two operators $P$ and $Q$ (which amounts
to a rescaling of the two currents $J$ and $\bar J$ in (\ref{curr})).

\par
When this last relation holds,  the currents obey the identity
\bea
\frac{1}{2}\left(Q-P\right)\left[g^{-1}\left(\dd \bar J +\bar\dd J\right)g\right]   
+\frac{1}{2}\left(Q+P\right)\left[g^{-1}\left(\dd \bar J -\bar\dd J 
+ \varepsilon\left[J\,,\,\bar J\right]\right)g\right] =0\,\,\,\,\,.
\eea
If in addition, the operator  $\left(Q+P\right)$ is invertible then the two 
currents $J$ and $\bar J$  obey the two relations 
\bea
\dd \bar J + \bar\dd J &=& 0\,\,\,\,,
\nonumber \\
\dd \bar J -\bar\dd J 
+ \varepsilon\left[J\,,\,\bar J\right] &=& 0\,\,\,\,.
\label{EOM}
\eea
Therefore, in addition of being on-shell conserved, the currents  $J$ and $\bar J$ 
have zero curvature.
\par
These last two equations are the consistency conditions of the linear differential system
\bea
\left\{
\begin{array}{l}
\left(\dd +\frac{\varepsilon}{1+\mu}\,J\right)\Psi= 0\\
\left(\bar \dd +\frac{\varepsilon}{1-\mu}\,\bar J\right)\Psi= 0
\end{array} \right. \,\,\,\,.
\label{lin-sys}
\eea
Here $\Psi\left(z,\bar z,\mu\right)$ is a matrix valued field. The requirement that this linear 
differential system 
is consistent,  for all values of the spectral parameter $\mu$, leads to the equations of motion 
of the non-linear sigma model (\ref{EOM}). This is preciseley the statement of the classical integrability
of a two-dimensional  non-linear sigma model \cite{zakharov}.

Finally, in terms of the linear operators $P$ and $Q$, the relation (\ref{bi-linear})
involving  the bi-linear form $<\,\,,\,\,>_{\cal G}$ becomes upon using (\ref{PQ--1})
\be
<X\,,\,Q^{-1} Y>_{{\cal G}} \,\,=\,\, <P^{-1}X\,,\,Y>_{{\cal G}} 
-2\lambda <X\,,\,Y>_{{\cal G}}\,\,\,\,.
\label{bi-linear-1}
\ee 
By writting $X=PZ$ and $Y=QW$, where $X$, $Y$, $Z$, and $W$ are in the Lie algebra 
${\cal G}$, this last relation becomes 
\be
<PZ\,,\,W>_{{\cal G}} \,\,=\,\, <Z\,,\,QW>_{{\cal G}} 
-2\lambda <PZ\,,\,QW>_{{\cal G}}\,\,\,\,.
\ee

\par
\noindent
{\bf Summary :}
\par
\noindent
Given two linear operators $P$ and $Q$ (we assume that $P$, $Q$ and $P+Q$ are invertible) on a Lie
algebra ${\cal G}$ and satisfying, for any two elements $X$ and $Y$ in ${\cal G}$,  the two
relations
\bea
&<PX\,,\,Y>_{{\cal G}} \,\,=\,\, <X\,,\,QY>_{{\cal G}} 
-2\lambda <PX\,,\,QY>_{{\cal G}}\,\,\,\,,&
\label{cond1}
\\
&\,&
\nonumber \\
&\left[PX \,,\, Q Y\right] - P\left[X \,,\, Q Y\right]- Q\left[PX \,,\,  Y\right]
=\frac{\varepsilon}{2}\left(P+Q\right)\left[X \,,\, Y\right]&
\label{cond2}
\eea
then the two-dimensional non-linear sigma model defined by the action
\bea
S\left(g\right) &=& \lambda  \int_{\dd{\cal{M}}}{\rm d}z{\rm d}\bar z \,
<g^{-1}\dd g\,\,,\,\,g^{-1}\bar\dd g>_{{\cal G}}
\nonumber \\
 &+&  \frac{\lambda}{6}\int_{{\cal{M}}}{\rm d}^3x\,\epsilon^{\mu\nu\rho}\,<g^{-1}\dd_\mu g\,\,,\,\,
\left[\,g^{-1}\dd_\nu g\,\,,\,\, g^{-1}\dd_\rho g\right]\,>_{{\cal G}}
\nonumber \\
&+&  \int_{\dd{\cal{M}}}{\rm d}z{\rm d}\bar z \,
<g^{-1}\dd g\,\,,\,\,
Q^{-1}\,\left(g^{-1}\bar\dd g\right)>_{{\cal G}}
\,\,\,\,\,
\label{sig-mod-fin}
\eea
is classically  integrable. We have used (\ref{PQ--1}) to write  $M+N=Q^{-1} +\lambda I$.
The equations of motion stemming from this action  are written in (\ref{EOM}) in terms 
of two the currents $J$ and  $\bar J$ 
\bea
J &=& g\left[P^{-1}\left(g^{-1}\dd g\right)\right]g^{-1}\,\,\,\,,
\nonumber \\
\bar J &=& g\left[Q^{-1}\left(g^{-1}\bar\dd g\right)\right]g^{-1} \,\,\,\,
\label{curr-1}
\eea
and are equivalent to  the consistency 
conditions of the linear system (\ref{lin-sys}).

\section{The Yang-Baxter sigma model}

The so-called Yang-Baxter non-linear sigma model is obtained 
as a special case of our construction. Indeed, 
let us first assume that the two linear operators are of the form
\bea
P &=& \kappa I + \zeta R\,\,\,\,\,,
\nonumber \\
Q &=& \kappa I - \zeta R\,\,\,\,\,,
\eea
where  $R$ is a linear operator
acting on the generators of the Lie algebra ${\cal G}$ and 
$\kappa$ and $\zeta^2=-\kappa\left(\kappa+\varepsilon\right)>0 $ are two constants. The parameters 
$\kappa$ and $\varepsilon$ are such  $\zeta^2$ is strictly positive. 
We also put the Wess-Zumino-Witten term in the action to zero. That is,
\be
\lambda=0\,\,\,\,.
\ee
When  $\zeta^2=-\kappa\left(\kappa+\varepsilon\right)$, the two relations in (\ref{cond1}) 
and (\ref{cond2}) become then respectively
\bea
 & <RX\,,\,Y>_{{\cal G}} + <RY\,,\,X>_{{\cal G}}=0\,\,\,\,,&
\nonumber \\ 
&\,&
\nonumber \\
& \left[RX \,,\, R Y\right] - R\left(\left[RX \,,\,  Y\right] + \left[X \,,\,  RY\right]\right)
=\left[X \,,\, Y\right]\,\,\,\,. &
\label{R-operator}
\eea 
The last relation is known as the modified Yang-Baxter equation while the firt
equation says that the linear operator $R$ is anti-symmetric with respect to the bi-linear form. 
A solution to these relations is  given in  \cite{jimbo,klimcik} and is briefly recalled in the next section.
\par
The corresponding action is obtained upon replacing $Q^{-1}$ in (\ref{sig-mod-fin}) and is  given by 
\bea
S\left(g\right) = \int_{\dd{\cal{M}}}{\rm d}z{\rm d}\bar z \,
<g^{-1}\dd g\,\,,\,\,
\left(\kappa I - \zeta R\right)^{-1}\left(g^{-1}\bar\dd g\right)>_{{\cal G}}
\,\,\,\,\,\,\,\,,\,\,\,\,\,\,\,\, \zeta^2=-\kappa\left(\kappa+\varepsilon\right) >0
\,\,\,.
\label{I}
\eea
This is precisely the action found in \cite{klimcik}.

\section{Constructing a solution }

Our main concern now is to find solutions to  (\ref{cond1}) and   (\ref{cond2}).
We start by recalling the commutation relations of a Lie algebra in the 
Cartan-Weyl basis
\bea
\left[H_i\,,\,H_j\right] &=& 0\,\,\,\,\,\,\,\,,\,\,\,\,\,\,\,\,i,j=1\dots r\,\,\,,
\nonumber \\
\left[H_i\,,\,E_\alpha\right] &=& \alpha_i E_\alpha\,\,\,,
\nonumber \\
\left[E_\alpha\,,\,E_{-\alpha}\right] &=&  \alpha_i H_i\,\,\,,
\nonumber \\
\left[E_\alpha\,,\,E_\beta\right] &=& 
\left\{
\begin{array}{ll}
{\cal  {N}}_{\alpha,\beta}E_{\alpha+\beta} & {\rm{if}}\,\,\,\,\,\,\alpha+\beta\in\Sigma\,\,\,, \\
0  & {\rm{if}}\,\,\,\,\,\,\alpha+\beta\notin\Sigma\,\,\,.
\end{array} \right.
\label{Cartan-Weyl}
\eea
Here $\Sigma$ is the set of roots\footnote{We use the conventions and notations of ref.\cite{michel}}. 
The generators are normalised such that
the Killing form (the bi-linear form) is
\bea
<H_i\,,\,H_j> = \delta_{ij} \,\,\,\,\,\,\,\,,\,\,\,\,\,\,\,\,
<H_i\,,\,E_\alpha> =0 \,\,\,\,\,\,\,\,,\,\,\,\,\,\,\,\,
<E_{\alpha}\,,\,E_\beta> =\delta_{\alpha+\beta,0} \,\,\,.
\label{Killing-form}
\eea
\par
Since we will use the linear operator $R$, defined in (\ref{R-operator}), 
we start by giving its action on the generators of the Lie algebra in the 
Cartan-Weyl basis as found in \cite{jimbo,klimcik}. This is 
\bea
\left\{
\begin{array}{ll}
R\,H_i = 0 \,\,\,,& \\
& \\
R\, E_\alpha = -i E_\alpha &  {\rm{if}}\,\,\,\,\,\,\alpha\in\Sigma^+\,\,\,, \\
& \\
R\, E_{-\alpha} = i E_{-\alpha} &  {\rm{if}}\,\,\,\,\,\,\alpha\in\Sigma^+\,\,\,, 
\end{array} \right.
\,\,\,\,
\label{Cartan-R}
\eea
where $\Sigma^+$ is the set of positive roots and $i^2=-1$ (not to be confused with the index $i$ used above).
The action of the linear operator $R$ on the generators of the Lie algebra in the basis
$\left\{T_a\right\}$ is specified by
\bea
&& \,\,\,\, R\,T_a=0 \,\,\,\, \textrm{if}\,\, \,\,T_a \in {\cal{H}}\,\,\,\,,
\nonumber \\
&&
\left.
\begin{array}{l}
R\,T_a = T_{a+1}\,\,\,\,, \\
\\
R\, T_{a+1}=-T_a \,\,\,\,,
\end{array}\right\}\,\,\,\, \textrm{with}\,\, \,\,E_{\alpha_{a}} = T_a+i T_{a+1} \,\,  \textrm{and such that } \,\,\,\alpha_a \in\Sigma^+ \,\,\,.
\label{R-action}
\eea
Here $ {\cal{H}} $ is the Cartan subalgebra of the Lie algebra  ${\cal{G}}$.
\par
It is instructive to illustrate the  action  of the linear operator $R$ on the generators of the Lie algebra $SU(3)$. 
The generalisation 
to other Lie algebras can be figured out  in a similar manner. 
The $SU(3)$ Cartan-Weyl basis is constituted as 
\be
E_{{\pm\alpha}_{(1)}} = T_1\pm iT_2\,\,\,,\,\,\,E_{{\pm\alpha}_{(2)}} = T_4\pm iT_5\,\,\,,\,\,\,E_{{\pm\alpha}_{(3)}} = T_6\pm iT_7\,\,\,,\,\,\,
H_1=T_3 \,\,\,,\,\,\,H_2=T_8 \,\,\,.
\ee
Using (\ref{R-action}), one finds that the operator $R$ acts on the $SU(3)$ generators $\{T_a\}$ as 
\bea
R
\left(
\begin{array}{l}
T_1\\T_2\\T_3\\T_4\\T_5\\T_6\\T_7\\T_8
\end{array} \right)=
\left(
\begin{array}{cccccccc}
0 & 1 &0 &  0&  0&  0&  0&  0 \\
-1 & 0  &0 &  0&  0&  0&  0&  0\\
0& 0& 0 & 0& 0& 0& 0& 0 \\
0& 0& 0& 0& 1 & 0& 0& 0 \\
0& 0& 0&-1 & 0 & 0& 0& 0 \\
0& 0& 0&  0& 0& 0 & 1 & 0 \\
0& 0& 0&  0& 0& -1 & 0 & 0 \\
0& 0& 0&  0& 0& 0& 0& 0
\end{array}\right)
\left(
\begin{array}{l}
T_1\\T_2\\T_3\\T_4\\T_5\\T_6\\T_7\\T_8 
\end{array} \right)\,\,\,.
\eea
It is then clear that the matrix $R^2$ is diagonal with entries equal to either $-1$
or $0$ (zero corresponds to the action of   $R^2$ on the elements of the Cartan subalgebra). 
The operator  $R^2$ will be needed later.

\par
Lut us now return to the linear operators $P$ and $Q$.  We assume that they act on the generators of the 
Lie algebra in the Cartan-Weyl basis  as
\bea
\left\{
\begin{array}{ll}
P\,H_i = \sigma_i H_i\,\,\,,& \\
& \\
P\,E_\alpha = pE_\alpha &  {\rm{if}}\,\,\,\,\,\,\alpha\in\Sigma^+\,\,\,, \\
& \\
P\,E_{-\alpha} = p^*E_{-\alpha} &  {\rm{if}}\,\,\,\,\,\,\alpha\in\Sigma^+\,\,\,, 
\end{array} \right.
\,\,\,\,\,\,\,\,,\,\,\,\,\,\,\,\,
\left\{
\begin{array}{ll}
Q\,H_i = \xi_i H_i\,\,\,,& \\
& \\
Q\,E_\alpha = qE_\alpha &  {\rm{if}}\,\,\,\,\,\,\alpha\in\Sigma^+\,\,\,, \\
& \\
Q\,E_{-\alpha} = q^*E_{-\alpha} &  {\rm{if}}\,\,\,\,\,\,\alpha\in\Sigma^+\,\,\,, 
\end{array} \right.
\label{PQ-cartan}
\eea
where no summation over the repeated
index $i$ is implied. The constants $\sigma_i$ and $\xi_i$ are real while
$p$ and $q$ are complex.
In the basis $\left(H_i\,,\,E_\alpha\,,\,E_{-\alpha}\right)$, the matrices associated to the operators $P$ and $Q$ 
are diagonal. 

\par

\par
Using the commutation relations (\ref{Cartan-Weyl}), the Killing form (\ref{Killing-form}) and the action of the linear operators as in 
(\ref{PQ-cartan}), the relations (\ref{cond1}) and   (\ref{cond2}) are satisfied if
\bea
-pq &=& \frac{\varepsilon}{2}\left(p+q\right)\,\,\,\,\,,
\label{rel-1}
\\
pq^* -q^*\sigma_i - p\xi_i &=& \frac{\varepsilon}{2}\left(\sigma_i+\xi_i\right)\,\,\,\,\,,
\label{rel-2}
\\
\sigma_i &=&\xi_i -2\lambda \sigma_i\xi_i \,\,\,\,\,,
\label{rel-3}
\\
p &=& q^* - 2\lambda pq^* \,\,\,\,\,.
\label{rel-4}
\eea
The last two equations give simply $\sigma_i$ in terms of  $\xi_i$  and $p$ in terms of $q$
\bea
\sigma_i &=& \frac{\xi_i}{1+2\lambda \xi_i}\,\,\,,
\label{sigma}
\\
p &=& \frac{q^*}{1+2\lambda q^*}\,\,\,.
\label{p}
\eea
\par
Upon reporting (\ref{sigma}) and (\ref{p}) in (\ref{rel-2}) and (\ref{rel-1}) one finds
\bea 
&& q^* =   \tau_j
 \pm \,i\sqrt{-\tau_j\left(\tau_j + \frac{\varepsilon}{1+\lambda\varepsilon}\right)}
\,\,\,\,\,\,,\,\,\,\,\,\,j=1\dots r
 \,\,\,,
\nonumber \\
&&  \left(1+\lambda\varepsilon\right)q^*q+ \frac{\varepsilon}{2}\left(q^*+q\right)=0\,\,\,. 
\label{q}
\eea
Here $i^2=-1$ and $\tau_j$ is defined as
\bea
\tau_j =  \xi_j\left(1+\lambda\xi_j\right)\left(1+\lambda\varepsilon\right) \,\,\,\,\,\,,\,\,\,\,\,\,j=1\dots r
\,\,\,.
\eea
We have therefore determined $q$ in terms of $\xi_j$. The two equations in (\ref{q}) are always compatible.
Next, the parameter $p$ is calculated from  (\ref{p}).
\par
Now two paramaters $\xi_i$ and $\xi_j$, say, must lead to the same value of $q$ according to (\ref{q}). This means that
we must have also
\be
\xi_i\left(1+\lambda\xi_i\right)\left(1+\lambda\varepsilon\right) =\xi_j\left(1+\lambda\xi_j\right) \left(1+\lambda\varepsilon\right)
\,\,\,\,\,\,\,\,,\,\,\,\,\,\,\,\,i,j=1\dots r\,\,\,.
\ee
Therefore, the parameters $\xi_i$ are such that
\bea
\xi_i =\xi_j 
\,\,\,\,\,\,\,\,{\mathrm {or}}\,\,\,\,\,\,\,\,
\xi_i = -\frac{1+\lambda\xi_j}{\lambda}
\,\,\,\,\,\,\,\,,\,\,\,\,\,\,\,\,i,j=1\dots r\,\,\,.
\label{choice-xi}
\eea
This means that they fall into two sets $\left\{\xi_1,\dots,\xi_{r-l}\right\}$
and $\left\{\xi_{r-l-1},\dots,\xi_{r}\right\}$, $ 0 \le l \le r $, and
the members of a  set are identical. A set could be empty (if $l=0$).
The corresponding expressions for the parameters  $\sigma_i$ are found from  (\ref{sigma}).

These two choices for the constants  $\xi_i$ 
suggests the splitting of the Cartan subalgebra of ${\cal G}$ as
\be
{\cal H} = {\cal H}_{r-l}\, \cup\, {\cal H}_l \,\,\,\,,\,\,\,\, 0 \le l \le r\,\,\,\,,
\ee
where $ {\cal H}_{r-l}  $ contains the first $r-l$ elements of ${\cal H}$ and 
${\cal H}_l$ the remaining  $l$ elements ($ 0 \le l \le r $).

\par
We have now all the ingredients to put forward the full solution to 
the equations  (\ref{rel-1})--(\ref{rel-4}).  This is given by  
\bea
&&  \xi_1 = \xi_2 = \dots = \xi_{r-l} =\xi \,\,\,, \,\,\,
 \xi_{r-l+1} = \xi_{r-l+2}  =\dots = \xi_{r} =  -\frac{1+\lambda\xi}{\lambda} \,\,\,, 
\nonumber \\
&& \sigma_1 = \sigma_2 =\dots = \sigma_{r-l}= \frac{\xi}{1+2\lambda \xi} \,\,\,,  \,\,\,
\sigma_{r-l+1} = \sigma_{r-l+2}  =\dots = \sigma_{r} =  \frac{1+\lambda\xi}{\lambda\left(1+2\lambda\xi\right)}  \,\,\,, 
\nonumber \\
&& q= \tau \,\mp\, i\, \omega \,\,\,, 
\nonumber \\
&& p= \tau^\prime \,\pm\, i\, \omega^\prime \,\,\,, 
\label{sol-4}
\eea
where   $\xi$ is  a  free parameter. The constants $\tau$, $\omega$, $\tau^\prime$ and  $\omega^\prime$  are given by
\bea
\tau &=&  \xi\left(1+\lambda\xi\right)\left(1+\lambda\varepsilon\right)\,\,\,,
\nonumber \\
\omega &= & \sqrt{-\tau\left(\tau + \frac{\varepsilon}{1+\lambda\varepsilon}\right)} \,\,\,,
\nonumber \\
\tau^\prime &=& \frac{\xi\left(1+\lambda\xi\right)\left(1-\lambda\varepsilon\right) }{\left(1+2\lambda\xi\right)^2} 
\,\,\,,
\nonumber \\
\omega^\prime &=& \frac{1} {\left(1+2\lambda\xi\right)^2} \sqrt{-\tau\left(\tau + \frac{\varepsilon}{1+\lambda\varepsilon}\right)} \,\,\,.
\label{omega}
\eea
The only restriction on the free parameter $\xi$ is that the argument of the square root
in the expression of $\omega$ is positive or zero. This is equivalent to demanding
that
\be
\tau = \xi\left(1+\lambda\xi\right)\left(1+\lambda\varepsilon\right)
\in\,\left[ -\frac{\varepsilon}{1+\lambda\varepsilon}\,,\,0\right[ \,\,\,.
\label{domain}
\ee
The domain of parameters is therefore  quite vast.

\section{The integrable non-linear sigma model}

The linear operators $P$ and $Q$ acting on the basis  $\{T_a\}$ 
of the Lie algebra ${\cal G}$ are deduced from (\ref{PQ-cartan}) and the solution (\ref{sol-4}).
It might be helpful to work out their action on the Lie algebra $SU(3)$ first.
For instance, $Q E_{\alpha_{(1)}}=q\,E_{\alpha_{(1)}}=\left(\tau \mp i\omega\right)E_{\alpha_{(1)}}$
implies that    $Q\,T_1 =\tau T_1  \pm \omega T_2$ and $Q\,T_2 =\tau T_2  \mp \omega T_1$,
and so on. If we partition the $SU(3)$ Cartan subalgebra as 
${\cal H} = {\cal H}_{r-l}\, \cup\, {\cal H}_l=T_3 \, \cup\, T_8$ then we have

\bea
Q
\left(
\begin{array}{l}
T_1\\T_2\\T_3\\T_4\\T_5\\T_6\\T_7\\T_8
\end{array} \right)=
\left(
\begin{array}{cccccccc}
\tau & \pm\omega &0 &  0&  0&  0&  0&  0 \\
\mp\omega & \tau  &0 &  0&  0&  0&  0&  0\\
0& 0& \xi & 0& 0& 0& 0& 0 \\
 0& 0& 0&\tau & \pm\omega & 0& 0& 0 \\
 0& 0& 0&\mp\omega & \tau  & 0& 0& 0 \\
0& 0& 0&  0& 0& \tau & \pm\omega & 0 \\
0& 0& 0&  0& 0& \mp\omega & \tau & 0 \\
0& 0& 0&  0& 0& 0& 0& -\frac{1+\lambda\xi}{\lambda}
\end{array}\right)
\left(
\begin{array}{l}
T_1\\T_2\\T_3\\T_4\\T_5\\T_6\\T_7\\T_8
\end{array} \right)\,.
\eea
The matrix corresponding to the operator $P$ can be determined in a similar manner.
\par
{} For the sake of condensing the expressions, we introduce the notation
\bea
\gamma &=& \xi -\tau  \,\,\,\, =  \,\,\,\,
-{\lambda \xi}\left[\varepsilon+\xi\left(1+\lambda\varepsilon\right) \right]
\,\,\,,\nonumber \\
\rho &=&   -\frac{1+\lambda\xi}{\lambda} -\tau \,\,\,\, =  \,\,\,\,
-\frac{\left(1+\lambda \xi\right)}{\lambda}\left[1+\lambda\xi\left(1+\lambda\varepsilon\right) \right]
\,\,\,,\nonumber \\
\gamma^\prime &=& \frac{\xi}{1+2\lambda \xi} -\tau^\prime  \,\,\,\,= \,\,\,\,
 \frac{\lambda \xi}{\left(1+2\lambda\xi\right)^2}\left[\varepsilon+\xi\left(1+\lambda\varepsilon\right) \right]
\,\,\,,
\nonumber \\
\rho^\prime &=&   \frac{1+\lambda\xi}{\lambda\left(1+2\lambda\xi\right)}  -\tau^\prime  \,\,\,\,= \,\,\,\,
\frac{\left(1+\lambda \xi\right)}{\lambda\left(1+2\lambda\xi\right)^2}\left[1+\lambda\xi\left(1+\lambda\varepsilon\right) \right]
\,\,\,.
\label{notation}
\eea
The operators  $P$ and $Q$ are given by
\bea
P &=& \tau^\prime \, I
\,\mp\,  \omega^\prime\, R  + \gamma^\prime \, {\cal Z}_{r-l} + \rho^\prime\, {\cal Z}_{l} \,\,\,,
\nonumber \\
Q &=& \tau \, I
\,\pm\,  \omega \, R  + \gamma \, {\cal Z}_{r-l} + \rho\, {\cal Z}_{l} \,\,\,.
\label{PQ}
\eea
The linear operator $R$ is still that in (\ref{R-action}), $I$ is the identity operator and 
the action of the  linear operators  ${\cal Z}_{r-l}$ and ${\cal Z}_l$ on  the basis $\{T_a\}$ is 
\be
\left\{
\begin{array}{l}
{\cal Z}_{r-l}\,T_a =
\,T_a \,\,\,\,\,\,\,\textrm{only if} \,\,\,\,\,\, T_a\in {\cal H}_{r-l} \,\,\,\,,\,\,\,\, 0 \le l \le r\,\,\,\,, \\
\\
{\cal Z}_l\,T_a = 
\,T_a \,\,\,\,\,\,\,\textrm{only if} \,\,\,\,\,\, T_a\in {\cal H}_l \,\,\,\,,\,\,\,\, 0 \le l \le r\,\,\,\,, \\
\\
{\cal Z}_{r-l}\,T_a \,=\,{\cal Z}_{l}\,T_a \,=\,  0 \,\,\,\,\,\,\,\textrm{otherwise} \,\,\,\,.
\end{array} \right.
\label{Z-Z}
\ee
The operators  ${\cal Z}_{r-l}$ and ${\cal Z}_l$ act only on the elements of  the Cartan subalgebra 
${\cal H} = {\cal H}_{r-l}\, \cup\, {\cal H}_l$  with  $0 \le l \le r$. 

\par
The next step in our construction is the computation of the inverses of the two
operators $P$ and $Q$. These are block diagonal matrices having either
$2\times 2$ or $1\times 1$ matrices along the diagonal and are easily inverted.
Indeed, we have
\bea
P^{-1} &=& \frac{\tau^\prime}{{\tau^\prime}^2 +{\omega^\prime}^2}\,I \pm \frac{\omega^\prime}{{\tau^\prime}^2 +{\omega^\prime}^2}\,R
+\left(\frac{1}{\tau^\prime +\gamma^\prime} -\frac{\tau^\prime}{{\tau^\prime}^2 +{\omega^\prime}^2}\right) {\cal Z}_{r-l}
\nonumber \\
&+& \left(\frac{1}{\tau^\prime +\rho^\prime} -\frac{\tau^\prime}{{\tau^\prime}^2 +{\omega^\prime}^2}\right) {\cal Z}_{l}
\,\,\,\,,
\nonumber \\
Q^{-1} &=& \frac{\tau}{\tau^2 +\omega^2}\,I \mp \frac{\omega}{\tau^2 +\omega^2}\,R
+\left(\frac{1}{\tau +\gamma} -\frac{\tau}{\tau^2 +\omega^2}\right) {\cal Z}_{r-l}
\nonumber \\
&+& \left(\frac{1}{\tau +\rho} -\frac{\tau}{\tau^2 +\omega^2}\right) {\cal Z}_{l}
\,\,\,\,.
\label{inverse-Q}
\eea
Explicitly, these expressions give
\bea
P^{-1} &=& -\frac{1}{\varepsilon}\left[\left(1-\lambda\varepsilon\right)I \,\pm\, \sqrt{-\alpha\beta}\, R  
+ \alpha\,{\cal Z}_{r-l} +\beta\,{\cal Z}_{l} \right] \,\,\,\,,
\nonumber \\
Q^{-1} &=&   -\frac{1}{\varepsilon}\left[\left(1+\lambda\varepsilon\right)I \,\mp\,  \sqrt{-\alpha\beta}\, R  
+ \alpha\,{\cal Z}_{r-l} + \beta\,{\cal Z}_{l} \right] \,\,\,\,.
\label{inverse-Q-a}
\eea
The two constants $\alpha$ and $\beta$ are defined as
\bea
\alpha = - \frac{1}{\xi}\left[\varepsilon+\xi\left(1+\lambda\varepsilon\right) \right]
\,\,\,\,\,\,\,\,\,\,,\,\,\,\,\,\,\,\,\,\,
\beta = - \frac{\left[1+\lambda\xi\left(1+\lambda\varepsilon\right) \right] }{\left(1+\lambda\xi\right)}
\,\,\,\,\,.
\label{alpha-beta}
\eea
By eliminating the parameter $\xi$ between  $\alpha$ and $\beta$, we find
that
\be
\beta=-\left[1-\frac{\left(\lambda\varepsilon \right)^2}{1+\alpha}\right]\,\,\,\,.
\label{beta}
\ee

\par
In terms of the parameters  $\alpha$ and $\beta$, the operators $P$ and $Q$ are as given in (\ref{PQ})
where
\bea
\tau &=& -\frac{\varepsilon\left(1+\alpha\right)\left(1+\lambda\varepsilon\right)}{\left(1+\alpha+\lambda\varepsilon\right)^2}
\,\,\,\,\,\,,\,\,\,\,\,\,
\tau^\prime =  -\frac{\varepsilon\left(1+\alpha\right)\left(1-\lambda\varepsilon\right)}{\left(1+\alpha-\lambda\varepsilon\right)^2}
\,\,\,\,,
\nonumber \\
\omega &=& \frac{\varepsilon\left(1+\alpha\right)\sqrt{-\alpha\beta}}{\left(1+\alpha+\lambda\varepsilon\right)^2}
\,\,\,\,\,\,,\,\,\,\,\,\,
\omega^\prime  =  \frac{\varepsilon\left(1+\alpha\right)\sqrt{-\alpha\beta}}{\left(1+\alpha-\lambda\varepsilon\right)^2}
\,\,\,\,,
\nonumber \\
\gamma &=& \frac{\lambda\varepsilon^2\alpha}{\left(1+\alpha+\lambda\varepsilon\right)^2}
\,\,\,\,\,\,,\,\,\,\,\,\,
\gamma^\prime = -\frac{\lambda\varepsilon^2\alpha}{\left(1+\alpha-\lambda\varepsilon\right)^2}
\,\,\,\,,
\nonumber \\
\rho &=& -\frac{\left(1+\alpha\right)\left[1+\alpha-\left(\lambda\varepsilon\right)^2\right]}{\lambda\left(1+\alpha+\lambda\varepsilon\right)^2}
\,\,\,\,\,\,,\,\,\,\,\,\,
\rho^\prime =  \frac{\left(1+\alpha\right)\left[1+\alpha-\left(\lambda\varepsilon\right)^2\right]}{\lambda\left(1+\alpha-\lambda\varepsilon\right)^2}
\,\,\,\,.
\eea
We notice that the parameters $(\tau^\prime\,,\,\omega^\prime\,,\,\gamma^\prime\,,\,\rho^\prime)$
are obtained from  $(\tau\,,\,\omega\,,\,\gamma\,,\,\rho)$ by the change $\lambda\longrightarrow -\lambda$.

\par
There is another way of writing  the operators $P^{-1}$ and $Q^{-1}$. Let  ${\cal Z}_{r}$  be the operator that acts as
\be
\left\{
\begin{array}{l}
{\cal Z}_{r}\,T_a =
\,T_a \,\,\,\,\,\,\,\textrm{only if} \,\,\,\,\,\, T_a\in {\cal H} \,\,\,, \\
\\
{\cal Z}_{r}\,T_a  \,=\,  0 \,\,\,\,\,\,\,\textrm{otherwise} \,\,\,\,.
\end{array} \right.
\label{Z-r}
\ee
That is,  ${\cal Z}_{r}$ acts on all the generator in the Cartan subalgebra  ${\cal H}$. It satisfies 
the  relation
\be
{\cal Z}_{r} =I+R^2 \,\,\,.
\label{Z-R-1}
\ee
Furthermore, it can be seen that
\bea
{\cal Z}_{r-l}={\cal Z}_{r}-{\cal Z}_{l}=\left(I+R^2\right)- {\cal Z}_{l} \,\,\,.
\label{Z-R-2}
\eea
Using this last relation, we can write the operators $P^{-1}$ and $Q^{-1}$ in the form
\bea
P^{-1} &=& -\frac{1}{\varepsilon}\left[\left(1-\lambda\varepsilon +\alpha\right)I \,\pm\, \sqrt{-\alpha\beta}\, R  
+ \alpha\,R^2 +\left(\beta-\alpha\right)\,{\cal Z}_{l} \right] \,\,\,\,,
\nonumber \\
Q^{-1} &=&   -\frac{1}{\varepsilon}\left[\left(1+\lambda\varepsilon +\alpha \right)I \,\mp\,  \sqrt{-\alpha\beta}\, R  
+ \alpha\,R^2 + \left(\beta-\alpha\right)\,{\cal Z}_{l} \right] \,\,\,\,.
\label{inverse-Q-b}
\eea
A word of caution is necessary here. The operators $P^{-1}$ and $Q^{-1}$ are not invertible if
either $\left(1-\lambda\varepsilon +\alpha\right)=-\left(\beta-\alpha\right)$ or 
$\left(1+\lambda\varepsilon +\alpha\right)=-\left(\beta-\alpha\right)$. In this case
the operators $\left(1-\lambda\varepsilon +\alpha\right)I \,
+ \alpha\,R^2 +\left(\beta-\alpha\right)\,{\cal Z}_{l} $ or  $\left(1+\lambda\varepsilon +\alpha\right)I \,
+ \alpha\,R^2 +\left(\beta-\alpha\right)\,{\cal Z}_{l} $ will have  zeros as entries along the diagonal whenever
acting on the generators in ${\cal H}_l$. The expression of  $\beta$ in (\ref{alpha-beta})
gives $(1+2\lambda\xi)=0$ and  $\lambda=0$ as solutions to $\left(1-\lambda\varepsilon\right)=- \beta$ and
 $\left(1+\lambda\varepsilon\right)=- \beta$.  These are precisely
the two situations which are not allowed as can be seen from the solution   (\ref{sol-4}). 

\par
Using the expression of $Q^{-1}$ in (\ref{inverse-Q-b}), 
our action (\ref{sig-mod-fin}) takes then the form
\bea
S_l\left(g\right) & =&  - \frac{1}{\varepsilon}\,
\int_{\dd{\cal{M}}}{\rm d}z{\rm d}\bar z \,
<g^{-1}\dd g\,\,,\,\,
\big[  \left(1 +\alpha \right)I \,\mp\,  \sqrt{-\alpha\beta}\, R  
+ \alpha\,R^2 
\nonumber \\
&+&  \left(\beta-\alpha\right)\,{\cal Z}_{l} \, \big] \left(g^{-1}\bar\dd g\right)>_{{\cal G}}
\nonumber \\
 &+&  \frac{\lambda}{6}\int_{{\cal{M}}}{\rm d}^3x\,\epsilon^{\mu\nu\rho}\,<g^{-1}\dd_\mu g\,\,,\,\,
\left[\,g^{-1}\dd_\nu g\,\,,\,\, g^{-1}\dd_\rho g\right]\,>_{{\cal G}}
\,\,\,\,\,.
\label{sig-mod-fin-3-a}
\eea
The parameters $\alpha$ and $\beta$ are related by (\ref{beta}) and $\lambda\varepsilon$
is another free parameter ($\frac{1}{\varepsilon}$ is an overall factor).
This is the main result of this paper.
The above two dimensional non-linear sigma model is integrable. The two current
$J = g\left[P^{-1}\left(g^{-1}\dd g\right)\right]g^{-1}$
and $\bar J = g\left[Q^{-1}\left(g^{-1}\bar\dd g\right)\right]g^{-1}$, with  $P^{-1}$ and $Q^{-1}$
as given in (\ref{inverse-Q-b}), are conserved and have a  vanishing curvature on-shell.
\par
At this stage a remark is due : In the case when $l=0$, that is when the set  ${\cal H}_l={\cal H}_0$ 
is an empty set (consequently  ${\cal Z}_0\,T_a=0$ for all $T_a$ in  the Lie algebra  ${\cal G}$), the 
action $S_0\left(g\right)$  is precisely that constructed in ref.\cite{delduc}. Their parameters, in this case, 
are related to ours as 
\be 
\eta^2=\alpha \,\,,\,\, A=\mp\sqrt{-\alpha\beta} \,\,,\,\,k^2=(\lambda\varepsilon)^2\,\,,\,\,
K=\frac{1}{\varepsilon}\,\,\,\,\,.
\label{diction}
\ee
With this identification, their relation $A=\eta\sqrt{1-\frac{k^2}{1+\eta^2}}$ is exactly that
written in  (\ref{beta}).
\par
In order to explore the novelty of our construction, we find it convenient to rewrite our final action as 
\bea
S_l\left(g\right) & =&  - \frac{1}{\varepsilon}\,
\int_{\dd{\cal{M}}}{\rm d}z{\rm d}\bar z \,
<g^{-1}\dd g\,\,,\,\,
\big[I +\alpha\,{\cal Z}_{r-l}+ \beta\,{\cal Z}_{l}  \,\mp\,  \sqrt{-\alpha\beta}\, R  
\, \big] \left(g^{-1}\bar\dd g\right)>_{{\cal G}}
\nonumber \\
 &+&  \frac{\lambda}{6}\int_{{\cal{M}}}{\rm d}^3x\,\epsilon^{\mu\nu\rho}\,<g^{-1}\dd_\mu g\,\,,\,\,
\left[\,g^{-1}\dd_\nu g\,\,,\,\, g^{-1}\dd_\rho g\right]\,>_{{\cal G}}
\,\,\,\,\,.
\label{sig-mod-fin-3-b}
\eea
In reaching this simplified version we have made use of (\ref{Z-R-1}) and  (\ref{Z-R-2}) and the action of 
the linear operators ${\cal Z}_{l}$ and ${\cal Z}_{r-l}$ is as defined in (\ref{Z-Z}).
The Cartan subalgebra is 
split as ${\cal H} = {\cal H}_{r-l}\, \cup\, {\cal H}_l$  with  $0 \le l \le r$ and ${\cal H}_0$ is the empty set.
\par
As mentioned above, the case $l=0$ is already treated in ref.\cite{delduc} and their 
integrable non-linear sigma model is given by the action
\bea
S_0\left(g\right) & =&  - \frac{1}{\varepsilon}\,
\int_{\dd{\cal{M}}}{\rm d}z{\rm d}\bar z \,
<g^{-1}\dd g\,\,,\,\,
\big[I  +\alpha\,{\cal Z}_{r}\,\mp\,  \sqrt{-\alpha\beta}\, R  
\, \big] \left(g^{-1}\bar\dd g\right)>_{{\cal G}}
\nonumber \\
 &+&  \frac{\lambda}{6}\int_{{\cal{M}}}{\rm d}^3x\,\epsilon^{\mu\nu\rho}\,<g^{-1}\dd_\mu g\,\,,\,\,
\left[\,g^{-1}\dd_\nu g\,\,,\,\, g^{-1}\dd_\rho g\right]\,>_{{\cal G}}
\,\,\,\,\,.
\label{delduc-et-al}
\eea
The linear operator ${\cal Z}_{r}=I+R^2$, given in (\ref{Z-r}), acts on all the generators in the Cartan subalgebra 
${\cal H}$.

\par
In the next section we will point out, by considering specific examples, that  the action  $S_l\left(g\right)$, for Lie algebras with rank $r\geq 2$,   contains 
deformations of the  Wess-Zumino-Witten model that are not accounted for by  the action $S_0\left(g\right)$
(the non-linear sigma model of ref.\cite{delduc}). Hence, this article is a  generalisation
of the work of  ref.\cite{delduc}.

\section{The deformed $SU\left(2\right)$ WZW model and beyond}

\par
It is instructive to illustrate our construction by first considering the Lie algebras 
$SU\left(2\right)$. {}For this purpose,  let us call 
\bea
{\cal D}_l = \alpha\,{\cal Z}_{r-l} +  \beta\,{\cal Z}_{l}  \,\mp\,  \sqrt{-\alpha\beta}\, R  
\eea
the deformation operator. 
We will also consider $\alpha$ and $\beta$ as our free parameters instead of 
$\alpha$ and $\lambda\varepsilon$. In terms of $\alpha$ and $\beta$, (\ref{beta}) gives
\bea
\left(\lambda\varepsilon \right)^2= \left(1+\alpha \right)\left(1+\beta \right)\,\,\,.
\eea
The deformed WZW action  (\ref{sig-mod-fin-3-b}) is then written as
\bea
S_l\left(g\right) & =&  - \frac{1}{\varepsilon}\,
\int_{\dd{\cal{M}}}{\rm d}z{\rm d}\bar z \,
<g^{-1}\dd g\,\,,\,\,
\big[I +{\cal D}_l
\, \big] \left(g^{-1}\bar\dd g\right)>_{{\cal G}}
\nonumber \\
 &+ & \frac{1}{6}\, \frac{1}{\varepsilon}\sqrt{\left(1+\alpha \right)\left(1+\beta \right)}
\int_{{\cal{M}}}{\rm d}^3x\,\epsilon^{\mu\nu\rho}\,<g^{-1}\dd_\mu g\,\,,\,\,
\left[\,g^{-1}\dd_\nu g\,\,,\,\, g^{-1}\dd_\rho g\right]\,>_{{\cal G}}
\,\,.
\label{sig-mod-fin-3-c}
\eea
Notice that the coefficient of the WZW term is symmetric under the exchange  $\alpha \leftrightarrow \beta $.

\par
In the case of the $SU\left(2\right)$ Lie algebra, with generators $\left\{T_1\,,\, T_2\,,\, T_3 \right\}$
and ${\cal H}= \left\{T_3 \right\}$, there are two deformation operators and their action is given by
\bea
{\cal D}_0
\left(
\begin{array}{l}
T_1\\T_2\\T_3
\end{array} \right) &=&
\left(
\begin{array}{ccc}
0 & \mp\sqrt{-\alpha\beta} &0  \\
\pm\sqrt{-\alpha\beta}  & 0  &0 \\
0& 0& \alpha
\end{array}\right)
\left(
\begin{array}{l}
T_1\\T_2\\T_3
\end{array} \right)\,\,\,\,.
\nonumber \\
\nonumber \\
{\cal D}_1
\left(
\begin{array}{l}
T_1\\T_2\\T_3
\end{array} \right) &=&
\left(
\begin{array}{ccc}
0 & \mp\sqrt{-\alpha\beta} &0  \\
 \pm\sqrt{-\alpha\beta} & 0  &0 \\
0& 0& \beta
\end{array}\right)
\left(
\begin{array}{l}
T_1\\T_2\\T_3
\end{array} \right)\,\,\,.
\eea
These  differ by their action on the generator $T_3$.
However, by the parameter  redefinition  $\alpha \leftrightarrow \beta $, the
deformation operators  ${\cal D}_0$ and ${\cal D}_1$ are mapped to each other\footnote {I thank an anonymous referee for
this remark.} and, therefore, lead to the same integrable non-linear sigma model.

\par
Despite the fact that we have established that the 
deformations operator  ${\cal D}_0$ and ${\cal D}_1$ are the same (up to a parameter redefinition), we will 
for completeness give the action for the deformed  $SU\left(2\right)$ WZW model.
The $SU\left(2\right)$ group element $g$ is parametrised as
\bea
g &=&
\left(
\begin{array}{cc}
\cos\left(\varphi_1\right)\,e^{-i\varphi_2} &  -\sin\left(\varphi_1\right)\,e^{-i\varphi_3} \\
\sin\left(\varphi_1\right)\,e^{i\varphi_3}   & \cos\left(\varphi_1\right)\,e^{i\varphi_2}
\end{array}\right) \,\,\,\,.
\eea
{}For the bi-linear form we take $<\,\,,\,\,>\,=\,\mathrm{Tr}$. 
The non-linear sigma model corresponding to the deformation operator ${\cal D}_0$
is given, up to a total derivative, by the action
\bea
S_0 & =&   \frac{2}{\varepsilon}\,
\int_{\dd{\cal{M}}}{\rm d}z{\rm d}\bar z \,\Big\{
\dd\varphi_1 \bar \dd\varphi_1 +\left[1+\alpha\cos^2(\varphi_1)\right]\cos^2(\varphi_1)\,
\dd\varphi_2 \bar \dd\varphi_2
\nonumber \\
 &+& \left[1+\alpha\sin^2(\varphi_1)\right]\sin^2(\varphi_1)\,
\dd\varphi_3 \bar \dd\varphi_3
- \alpha\cos^2(\varphi_1)\sin^2(\varphi_1)\left(\dd\varphi_2 \bar \dd\varphi_3 +\dd\varphi_3 \bar \dd\varphi_2\right)
\nonumber \\
&-& 2\sqrt{\left(1+\alpha \right)\left(1+\beta \right)} \,
\sin^2(\varphi_2-\varphi_3)\cos^2(\varphi_1)\left(\dd\varphi_2 \bar \dd\varphi_3 -\dd\varphi_3 \bar \dd\varphi_2\right)
\Big\}
\,\,.
\eea
The deformation operator ${\cal D}_1$ yields the same action with the replacement 
$\alpha\longrightarrow \beta$.

\par
Next, we consider the Lie algebra  $SU(3)$. Its  Cartan subalgebra is ${\cal H}=\left\{T_3\,,\,T_8\right\}$. 
The three deformation operators are
\bea
{\cal D}_0
\left(
\begin{array}{l}
T_1\\T_2\\T_3\\T_4\\T_5\\T_6\\T_7\\T_8
\end{array} \right) &=&
\left(
\begin{array}{cccccccc}
0 &A &0 &  0&  0&  0&  0&  0 \\
-A  & 0  &  0  &  0&  0&  0&  0&  0\\
0& 0 & \alpha & 0 &0  & 0& 0& 0 \\
 0& 0&  0  & 0 & A  & 0& 0& 0 \\
 0& 0& 0&-A & 0  & 0 & 0& 0 \\
0& 0& 0&  0& 0& 0 & A & 0 \\
0& 0& 0&  0& 0& -A & 0 & 0 \\
0& 0& 0&  0& 0& 0& 0&  \alpha
\end{array}\right)
\left(
\begin{array}{l}
T_1\\T_2\\T_3\\T_4\\T_5\\T_6\\T_7\\T_8
\end{array} \right)\,\,\,\,,
\nonumber\\
\nonumber\\
{\cal D}_1
\left(
\begin{array}{l}
T_1\\T_2\\T_3\\T_4\\T_5\\T_6\\T_7\\T_8
\end{array} \right) &=&
\left(
\begin{array}{cccccccc}
0 &A &0 &  0&  0&  0&  0&  0 \\
-A  & 0  &  0  &  0&  0&  0&  0&  0\\
0& 0 & \alpha & 0 &0  & 0& 0& 0 \\
 0& 0&  0  & 0 & A  & 0& 0& 0 \\
 0& 0& 0&-A & 0  & 0 & 0& 0 \\
0& 0& 0&  0& 0& 0 & A & 0 \\
0& 0& 0&  0& 0& -A & 0 & 0 \\
0& 0& 0&  0& 0& 0& 0&  \beta
\end{array}\right)
\left(
\begin{array}{l}
T_1\\T_2\\T_3\\T_4\\T_5\\T_6\\T_7\\T_8
\end{array} \right)\,\,\,\,\,,
\nonumber\\
\nonumber\\
{\cal D}_2
\left(
\begin{array}{l}
T_1\\T_2\\T_3\\T_4\\T_5\\T_6\\T_7\\T_8
\end{array} \right) &=&
\left(
\begin{array}{cccccccc}
0 & A &0 &  0&  0&  0&  0&  0 \\
 -A  & 0  &  0  &  0&  0&  0&  0&  0\\
0& 0 & \beta & 0 &0  & 0& 0& 0 \\
 0& 0&  0  & 0 & A  & 0& 0& 0 \\
 0& 0& 0& -A & 0  & 0 & 0& 0 \\
0& 0& 0&  0& -0& 0 & A & 0 \\
0& 0& 0&  0& 0& -A& 0 & 0 \\
0& 0& 0&  0& 0& 0& 0 &  \beta
\end{array}\right)
\left(
\begin{array}{l}
T_1\\T_2\\T_3\\T_4\\T_5\\T_6\\T_7\\T_8
\end{array} \right)\,\,\,\,\,,
\eea
where $A=\mp\sqrt{-\alpha\beta}$
as in the dictionary (\ref{diction}). We see that 
${\cal D}_2$ and ${\cal D}_0$ are related by the parameter redefinition 
$\alpha\leftrightarrow \beta$. However, ${\cal D}_1$ and ${\cal D}_0$
cannot be related by any parameter redefinition. It seems, therefore, that there are two
independent deformations of the $SU(3)$ WZW model, namely $S_0\left(g\right)$ and $S_1\left(g\right)$.
This remains though to be verified by an explicit calculation.

\par
In general, one may decompose the Maurer-Cartan one-form along the Cartan-Weyl basis (\ref{Cartan-Weyl}) as 
\bea
g^{-1}dg=\left[e^\gamma_a E_\gamma + e^{-\gamma}_a E_{-\gamma} + 
e^{i_{(r-l)}}_a H_{i_{(r-l)}} + e^{i_{(l)}}_a H_{i_{(l)}}\right]d\varphi^a\, \,\,\,\,.
\eea
Here $\varphi^a(z,\bar z)$ are the $n$ local fields and the index $\gamma$ runs over the positive roots $\Sigma^+$.
The Cartan subalgebra is 
partitioned as ${\cal H} = {\cal H}_{r-l}\, \cup\, {\cal H}_l$  with  $0 \le l \le r$. 
The indices $i_{(r-l)}=0,\dots, l$
and $i_{(l)}=l+1,\dots, r$ are such that $H_{i_{(r-l)}}\in {\cal H}_{r-l}$ and  $H_{i_{(l)}}\in {\cal H}_{l}$.
The vielbiens are functions of $\varphi^a(z,\bar z)$ and $e^{i_{(0)}}_a=0$.
\par
Using the bi-linear form $<\,,\,>$ as given in (\ref{Killing-form}) and the action of the operator $R$
in (\ref{Cartan-R}) together with the action of the operators ${\cal Z}_{r-l}$  and ${\cal Z}_{l}$ as
defined in  (\ref{Z-Z}), we find that
\bea
S_l\left(g\right) & =&  - \frac{1}{\varepsilon}\,
\int_{\dd{\cal{M}}}{\rm d}z{\rm d}\bar z \,
\left[e^\gamma_a e^{-\gamma}_b + e^{-\gamma}_a e^{\gamma}_b +\left(1+\alpha\right)e^{i_{(r-l)}}_ae^{i_{(r-l)}}_b
+\left(1+\beta\right)  e^{i_{(l)}}_a  e^{i_{(l)}}_b\right]\dd\varphi^a\bar\dd\varphi^b
\nonumber \\
   &\pm& i \frac{\sqrt{-\alpha\beta}}{\varepsilon}\,
\int_{\dd{\cal{M}}}{\rm d}z{\rm d}\bar z \,
\left[e^\gamma_a e^{-\gamma}_b  - e^{-\gamma}_a e^{\gamma}_b \right]\dd\varphi^a\bar\dd\varphi^b
\nonumber \\
 &+ & \frac{1}{6}\, \frac{1}{\varepsilon}\sqrt{\left(1+\alpha \right)\left(1+\beta \right)} 
\int_{{\cal{M}}}{\rm d}^3x\,\epsilon^{\mu\nu\rho}\,<g^{-1}\dd_\mu g\,\,,\,\,
\left[\,g^{-1}\dd_\nu g\,\,,\,\, g^{-1}\dd_\rho g\right]\,>_{{\cal G}}
\,\,\,\,\,.
\label{sig-mod-fin-3-d}
\eea
We see that the non-linear sigma models defined by  $S_l\left(g\right)$,  $l=0\,\dots\,r $, share
the same anti-symmetric tensor field (coming from the last two terms) but differ by their target space metric
(coming from the first term). It is clear that the two models $S_0\left(g\right)$ and $S_r\left(g\right)$ are related by the 
parameter redefinition $\alpha\leftrightarrow \beta$. Apart from this, we are inclined to conjecture that
there are $r$ different integrable models given by $S_l\left(g\right)$, $l=0\,\dots\,r-1$.

\section{Conclusions and outlook}

We have presented in this work an integrable two-dimensional non-linear sigma model. It is 
a two-parameter deformation of the Wess-Zumino-Witten model. 
We have found a simple solution to the main integrability equations (\ref{cond1}) and  (\ref{cond2})
of this article. It remains to see if these relations admit other solutions.
The renormalisability of the sigma model studied here and its  possible connection
to string theories is another interesting subject to be explored. 

\par
There is a strong link between integrability and gauging as shown in \cite{sfetsos,hollowood}.
This property is not very neat here. Indeed, the general WZW model  (\ref{sig-mod-fin})
is related to another theory as follows: The 
non-linear sigma model as defined by the action
\bea
S\left(g\,,\,h\right) &=& \lambda  \int_{\dd{\cal{M}}}{\rm d}z{\rm d}\bar z \,
<g^{-1}\dd g\,\,,\,\,g^{-1}\bar\dd g>_{{\cal G}}
\nonumber \\
 &+&  \frac{\lambda}{6}\int_{{\cal{M}}}{\rm d}^3x\,\epsilon^{\mu\nu\rho}\,<g^{-1}\dd_\mu g\,\,,\,\,
\left[\,g^{-1}\dd_\nu g\,\,,\,\, g^{-1}\dd_\rho g\right]\,>_{{\cal G}}
\nonumber \\
&-&  \int_{\dd{\cal{M}}}{\rm d}z{\rm d}\bar z \,
< h^{-1}\dd  h  \,\,,\,\,
Q\,\left( h^{-1}\bar\dd  h\right) >_{{\cal G}}
\nonumber \\
&+&  \int_{\dd{\cal{M}}}{\rm d}z{\rm d}\bar z \,
\left(< h^{-1}\dd  h\,\,,\,\,g^{-1}\bar\dd g >_{{\cal G}} +  
      < h^{-1}\bar\dd  h\,\,,\,\,g^{-1}\dd g >_{{\cal G}}\right)
\,\,\,\,\,
\label{sig-mod-inter}
\eea
is invariant under the constant left multiplication $ h \longrightarrow l\,h$.
This can be gauged by introducing a two components gauge field  $B_\mu $, with   $\mu=z\,,\bar z$, 
transforming as  $B_\mu \longrightarrow lB_\mu l^{-1} -\dd_\mu l  l^{-1}$. The gauging is carried out by replacing
$ h^{-1}\dd_\mu  h$ with  $ h^{-1}\left(\dd_\mu + B_\mu \right) h$.
The choice of the gauge $ h=1$ leads then, after the use of (\ref{bi-linear-1}),  to the action 
\bea
S\left(g\,,\,B_\mu\right) &=& \lambda  \int_{\dd{\cal{M}}}{\rm d}z{\rm d}\bar z \,
<g^{-1}\dd g\,\,,\,\,g^{-1}\bar\dd g>_{{\cal G}}
\nonumber \\
 &+&  \frac{\lambda}{6}\int_{{\cal{M}}}{\rm d}^3x\,\epsilon^{\mu\nu\rho}\,<g^{-1}\dd_\mu g\,\,,\,\,
\left[\,g^{-1}\dd_\nu g\,\,,\,\, g^{-1}\dd_\rho g\right]\,>_{{\cal G}}
\nonumber \\
 &+&  \int_{\dd{\cal{M}}}{\rm d}z{\rm d}\bar z \,
<g^{-1}\dd g\,\,,\,\,
Q^{-1}\,\left(g^{-1}\bar\dd g\right)>_{{\cal G}}
\nonumber \\
&-&  \int_{\dd{\cal{M}}}{\rm d}z{\rm d}\bar z \,
<B-\left(P^{-1}-2\lambda I\right)\left(g^{-1}\dd g\right) \,\,,\,\,
Q\left[\bar B -Q^{-1}\left(g^{-1}\bar\dd g\right)\right] >_{{\cal G}}\,\,\,.
\nonumber\\
&&
\label{sig-mod-inter-2}
\eea
The equations of motion of the non-dynamical fields $B$ and $\bar B$ are $B=\left(P^{-1}-2\lambda I\right) \left(g^{-1}\dd g\right)$
and $\bar B =Q^{-1}\left(g^{-1}\bar\dd g\right)$. Substituting these into (\ref{sig-mod-inter-2}) 
we recover our  general WZW action  (\ref{sig-mod-fin}). 
\par
Now, the equations of motion corresponding to the original action  (\ref{sig-mod-inter}) are 
\bea
\dd\left[g\bar ag^{-1}\right]+\bar \dd\left[g\left(a +
2\lambda\,A \right)g^{-1}\right] &=& 0\,\,\,,
\nonumber \\
\dd\left[h\left(Q\,\bar a -\bar A\right)h^{-1} \right] +
\bar\dd \left[h\left(\left(P^{-1}-2\lambda I\right)^{-1} a - A\right)h^{-1}\right]
 &=& 0\,\,\,,
\eea
where $a=h^{-1}\dd h$ and $\bar a = h^{-1}\bar\dd h$ and $A$ and $\bar A$ are as defined in (\ref{gauge-conn}).
The operator $\left(P^{-1}-2\lambda I\right) $ is obtained from the expression of 
$P^{-1}$ by simply changing $\lambda$ to $-\lambda$ as can be seen from (\ref{PQ--1}).
These equations of motion, assuming that $P$ and $Q$ obey (\ref{cond1}) and  (\ref{cond2}),  
do not seem to derive from some zero curvature conditions. Yet, the 
gauge fixed action  (\ref{sig-mod-inter-2}) leads to integrable non-linear sigma model. This issue deserves 
to be investigated. As a matter of fact, this remark is true for most of the  integrable sigma models found in the 
literature.

\noindent
\vskip1.0cm
\noindent
{\bf{{Note added:}}} After the completion of this work we became aware of the existence of ref.\cite{lacroix}
where (\ref{cond2}) was also established.  
Their construction makes the formulations in \cite{delduc, borsato} more compact 
and is inspired by the works of  Klim\v{c}\'{\i}k  \cite{klimcik1,klimcik2,klimcik3}. 
Their assumption on the anti-symmetric operator $R$ is that it solves the homogeneous or inhomogeneous classical Yang-Baxter equation.
In the case  of the usual Drinfel'd-Jimbo solution, the $R$ matrix satisfies the important relation $R^3=-R$. They showed, in this particular case, 
that their integrable non-linear sigma model is precisely that found in \cite{delduc} (see their section 3.2).
Since our $R$ matrix obeys also  $R^3=-R$, we conjecture that our models with $l=1\,,\dots\,,r-1$ are not covered by the construction of  ref.\cite{lacroix}.

\end{document}